\documentclass{article}

\usepackage{amssymb}
\usepackage[dvips]{graphicx}

\newcommand{\R}{{\mathbb R}}
\newcommand{\BS}{{\mathbb S}}
\newcommand{\ve}{\varepsilon}
\newcommand{\cD}{\cal D}
\newcommand{\pa}{\partial}

\begin{document}

\title{Weak asymptotics method}

\author{V.~G.~Danilov\thanks{This work  was supported by the
Russian Foundation for Basic Research under grant No.~99-01-01074.}\\ \\
Moscow, Russia, pm@miem.edu.ru}

\date{}

\maketitle

\begin{abstract}
We present a new method for constructing solutions 
to nonlinear evolutionary equations describing 
the propagation and interaction of nonlinear waves.
\end{abstract}

In the present paper, with the help of some simple examples, 
I demonstrate a new approach to the construction of asymptotic
solutions to differential equations.
I choose very simple examples (the Hopf equation and its
multidimensional analog) and, on purpose,  
omit details in the proofs of estimates (which are obvious here). 

Usually, by saying that a function is an asymptotic
(approximate) solution of a differential equation, 
we mean that this function satisfies the equation 
with a small discrepancy. 
The smallness of the discrepancy is understood 
as the smallness in some uniform metric under the assumption that 
a small parameter tends to zero. 

A function is called a {\em weak asymptotic solution} 
if, after the substitution of this function into the equation, 
there is a discrepancy that is small in the weak sense  
as a small parameter tends to zero. 
In this case the functionals are assumed to depend on time as on
a parameter. 

For example, under this approach, 
the $C^\infty$-approximation of a generalized function 
turns out to be its weak asymptotics and we can choose
generalized functions to be the initial conditions
and use their approximations for constructing the solutions. 
In this case, we obtain a small parameter, 
which is either the parameter of approximation 
or the small parameter in the original equation.
In the latter case, this original small parameter is taken to be
the parameter of  approximation.

In fact, this approach is close to the ideas proposed by
J.~F.~Colombeau and other authors who constructed  
different algebras of generalized functions.
The method itself was first introduced in our papers 
with V.~M.~Shelkovich initiated by the works of J.~F.~Colombeau and  
M.~Oberguggenberger with coautors and by discussions 
with J.~A.~Marti, V.~V.~Zharinov, and S.~Pilipovich.
The difference with the traditional approach  
is that in our approach 
the mollifier is chosen not from the consideration of the
algebraic construction  
but from the consideration of the original differential
equation. 

In some cases (shock waves), 
the solution is independent of the choice of the mollifier,
while in other cases (solitons, kinks) the solution depends on
this choice.

If the original equation contains a small parameter, 
then we, in fact, deal with regularizations 
by small viscosity or small dispersion. 
In this case, to calculate a weak asymptotics, 
we need to calculate the zero viscosity or zero dispersion
limits. 
Hence we arrive at the problem of constructing a definition of a
weak solution which admits this passage to the limit. 
 
It turned out that the approach developed here 
can be used for describing both the propagation of nonlinear
waves and, which is the most important, their interaction. 
It is well known that the problem of interaction 
of nonlinear traveling waves (for instance, of two kinks or two
solitons)  
in the case of a single spatial variable 
can be formulated as a problem of constructing 
the exact solution of a nonlinear equation 
with two spatial variables corresponding to the coordinates 
of the wave fronts. 
If the initial problem can be integrated 
by the method of the inverse scattering problem 
or in any other way, 
then one can write out the solution of the above-mentioned
equation explicitly,
which allows one to describe the interaction analytically. 
In our approach, to describe the interaction, one needs 
to solve an ordinary differential equation 
(or a system of such equations)
with a small parameter. Solutions of such equations can be
constructed by using a well-known technique. 
In what follows, 
we consider the main technical tools and some examples which
allow us to demonstrate the abilities of our approach.

\section{Some weak asymptotic formulas}

{\bf(a)} Let $\omega(z)\in \BS(\R^1)$, where $\BS$ is the Schwartz
space. We consider the function $\frac1\ve\omega((x-a)/\ve)$ and
calculate its weak asymptotics.
Treating $\frac1\ve\omega((x-a)/\ve)$ as a generalized function,
for any function $\eta(x)\in C^\infty_0$ 
we have
\begin{eqnarray*}
\langle\frac1\ve\omega\Big(\frac{x-a}\ve\Big),\eta(x)\rangle
&=&\frac1\ve\int\omega\Big(\frac{x-a}\ve\Big)
\eta(x)\,dx
=\int\omega(z)\eta(a+\ve z)\,dz\\
&=&\sum_{k\geq0}\Omega_k\frac{\ve^k}{k!}(-1)^k
\langle\delta^{(k)}(x-a),\eta\rangle, \qquad 
\ve>0,
\end{eqnarray*}
where the last relation is formal and means 
that the left-hand side can be represented as the asymptotic series
given on the right-hand side, 
$$
\Omega_k=\int\omega(z)z^k\,dz.
$$
We define by $O_{\cD'}(\ve^\alpha)$ and element of $\cD'$ 
such that 
$$
f(x,\ve)=O_{\cD'}(\ve^\alpha)\Leftrightarrow
\langle f(x,\ve),\eta(x)\rangle=O(\ve^\alpha),
$$
where the last $O$-estimate 
(which must hold for any function $\eta(x)\in C^\infty_0$)
is understood in the usual sense.
Then for any $N$ we can write 
$$
\frac1\ve\omega\Big(\frac{x-a}{\ve}\Big)
=\sum^{N}_{k=0}\Omega_k\frac{\ve^k}{k!}(-1)^k\delta^{(k)}(x-a)
+O_{\cD'}(\ve^{N+1}).
$$

{\bf(b)} Let $\omega_1(z),\omega_2(z)\in \BS(\R^1)$.
Let us consider the weak asymptotics of the product 
$\omega_1((x-a_1)/\ve)\omega_2((x-a_2)/\ve)$.
We have
\begin{eqnarray*}
\langle \omega_1\Big(\frac{x-a_1}\ve\Big)
\omega_2\Big(\frac{x-a_2}\ve\Big),
\eta(x)\rangle
&=&\int\omega_1\Big(\frac{x-a_1}\ve\Big)
\omega_2\Big(\frac{x-a_2}\ve\Big)\eta(x)\,dx\\
&=&\ve\eta(a_1)\int\omega_1(z)
\omega_2\Big(z-\frac{\Delta a}\ve\Big)\,dz+O(\ve^2)\\
&=&\ve\eta(a_2)\int
\omega_1\Big(z+\frac{\Delta a}\ve\Big)\omega_2(z)\,dz+O(\ve^2),\\
\Delta a&=&a_2-a_1.
\end{eqnarray*}
Finally, we obtain the following formula that is uniform
and symmetric  in $a_1,a_2$:
\begin{equation}
\omega_1\Big(\frac{x-a_1}\ve\Big)\omega_2\Big(\frac{x-a_2}\ve\Big)
=\frac12[\ve\delta(x-a_1)+\ve\delta(x-a_2)]B\Big(\frac{\Delta a}\ve\Big)
+O_{\cD'}(\ve^2),
\label{1}
\end{equation}
where
$$
B\Big(\frac{\Delta a}\ve\Big)=\int\omega_1(z)
\omega_2\Big(z-\frac{\Delta a}\ve\Big)\,dz
=\int\omega_1\Big(z+\frac{\Delta a}\ve\Big)\omega_2(z)\,dz.
$$

(c) Now let $\omega_1(z),\omega_2(z)\in C^\infty$, 
$\frac{d\omega_i}{dz}\in\BS(\R^1)$, 
$\lim_{z\to-\infty}\omega_i=0$, $\lim_{z\to\infty}\omega_i=1$, 
$i=1,2$.

Let us calculate the weak asymptotics of the derivative
\begin{eqnarray*}
&&\frac{d}{dx}\omega_1\Big(\frac{x-a_1}\ve\Big)
\omega_2\Big(\frac{x-a_2}\ve\Big)\\
&&\qquad 
=\frac1\ve
\dot\omega_1\Big(\frac{x-a_1}\ve\Big)\omega_2\Big(\frac{x-a_2}\ve\Big)
+\frac1\ve
\omega_1\Big(\frac{x-a_1}\ve\Big)\dot\omega_2\Big(\frac{x-a_2}\ve\Big).
\end{eqnarray*}
Just as previously, we have
\begin{eqnarray*}
&&\frac1\ve{\dot\omega_1}\Big(\frac{x-a_1}\ve\Big)\omega_2\Big(\frac{x-a_2}\ve\Big)
+\frac1\ve
\omega_1\Big(\frac{x-a_1}\ve\Big)\dot\omega_2\Big(\frac{x-a_2}\ve\Big)\\
&&\qquad
=\delta(x-a_1)B_1\Big(\frac{\Delta a}{\ve}\Big)
+\delta(x-a_2)B_2\Big(\frac{\Delta a}{\ve}\Big)
+O_{\cD'}(\ve),
\end{eqnarray*}
where
$$
B_1\Big(\frac{\Delta a}{\ve}\Big)=\int\dot\omega_1(z)
\omega_2\Big(z-\frac{\Delta a}{\ve}\Big)\,dz,
\qquad
B_2\Big(\frac{\Delta a}{\ve}\Big)=\int
\omega_1\Big(z+\frac{\Delta a}{\ve}\Big)\dot\omega_2(z)\,dz,
$$
We have 
$$
B_1(\infty)=0,\quad B_1(-\infty)=1,\quad 
B_1(z)+B_2(z)\equiv1.
$$
Calculating the primitive, we obtain
\begin{equation}
\omega_1\Big(\frac{x-a_1}\ve\Big)\omega_2\Big(\frac{x-a_2}\ve\Big)
=\theta(x-a_1)B_1\Big(\frac{\Delta a}{\ve}\Big)
+\theta(x-a_2)B_2\Big(\frac{\Delta a}{\ve}\Big)
+O_{\cD'}(\ve).
\label{2}
\end{equation}

{\bf(d)}  Under the assumptions of item~{\bf(b)} and the condition 
that $\int\omega_i(z)\,dz=1$, the functions 
$\omega_i((x-a_i)/\ve)$ are approximations (weak asymptotics) 
of the functions $\ve\delta(x-a_i)$,
$$
\omega_i((x-a_i)/\ve)=\ve\delta_{\ve,i}(x-a_i)
$$
Hence we can rewrite (1) as
$$
\ve\delta_{\ve,1}(x-a_1)\ve\delta_{\ve,2}(x-a_2)
=\frac12[\ve\delta(x-a_1)+\ve\delta(x-a_2)]
B(\Delta a/\ve)
+O_{\cD'}(\ve^2).
$$
In a similar way, under the assumptions of item~{\rm(c)}, 
$\omega_i((x-a_1)/\ve)=\theta_{\ve,i}(x-a_i)$ are approximations
of the Heaviside $\theta$-function.  
Hence we can rewrite (2) as
$$
\theta_{\ve,1}(x-a_1)\theta_{\ve,2}(x-a_2)
=\theta(x-a_1)B_1(\Delta a/\ve)
+\theta(x-a_2)B_2(\Delta a/\ve)
+O_{\cD'}(\ve).
$$

\section{Nonlinear structures}

We show how the above formulas can be used to describe 
interaction of nonlinear structures.

{\bf(a)} Interaction of shock waves for the Hopf equation. 
Let us consider the Cauchy problem
$$
L[u]=u_t+(u^2)_x=0,\qquad 
u\big|_{t=0}=u_0+u_1\theta(-x+a_1)+u_2\theta(-x+a_2),
$$
where $u_i$ are positive constants, $a_2<a_1$.
We approximate the initial condition according to the formulas
from item~{\bf1\,(c)} and seek the weak asymptotics of the solution 
in the form
$$
u_\ve(x,t)=u_0+u_1\theta_{\ve,1}(-x+\varphi_1(t,\ve))
+u_2\theta_{\ve,2}(-x+\varphi_2(t,\ve)),
$$
$$
\varphi_1(0)=a_1,\qquad \varphi_2(0)=a_2.
$$
Calculating the weak asymptotics of the expression 
$(u_\ve)^2$ according to the formulas from item~1\,(c), 
we obtain
\begin{eqnarray}
(u_\ve)^2&=&u^2_0+(u^2_1+2u_0u_1)\theta(-x+\varphi_1)
+u^2_0+(u^2_2+2u_0u_2)\theta(-x+\varphi_2)\label{3}\\
&&
+2u_1u_2\bigg[\theta(-x+\varphi_1)B_1\Big(\frac{\Delta \varphi}\ve\Big)
+\theta(-x+\varphi_2)B_2\Big(\frac{\Delta \varphi}\ve\Big)\bigg]
+O_{\cD'}(\ve),
\nonumber
\end{eqnarray}
where
\begin{eqnarray*}
B_1(\Delta \varphi/\ve)
&=&\int\dot\omega_1(z)\omega_2(z+\Delta\varphi/\ve)\,dz,\\
B_2(\Delta \varphi/\ve)
&=&\int\omega_1(z-\Delta\varphi/\ve)\dot\omega_2(z)\,dz,\quad
\Delta\varphi=\varphi_2-\varphi_1,
\end{eqnarray*}
and, in contrast to item~{\bf1\,(c)}, 
we have $B_1(-\infty)=0$, 
$B_1(\infty)=1$, but as before, $B_1+B_2\equiv1$.

We substitute the approximation of $u_\ve(x,t)$ into the
equation and require that the relation 
$L[u_\ve]=O_{\cD'}(\ve)$ must be satisfied 
(this is the {\em definition of the weak asymptotics solution in
this case\/}). Moreover, {\em the function $L[u_\ve]$ must be weakly
piecewise continuous with respect to $t$ for each fixed~$\ve$}.  
We obtain
$$
L[u_\ve]=\sum^{2}_{k=1}\Big[u_k\frac{d\varphi_k}{dt}
-2u_0u_k-u^2_k-2u_1u_2B_k\Big(\frac{\Delta\varphi}{\ve}\Big)\Big]
\delta(-x+\varphi_k)
+O_{\cD'}(\ve).
$$
Hence, in view of the definition of the weak solution, we have
\begin{equation}
\frac{d\varphi_k}{dt}=2u_0+u_k+2u_{3-k}
B_k\Big(\frac{\Delta\varphi}{\ve}\Big),\quad k=1,2.
\label{4}
\end{equation}

For $\Delta\varphi<0$ (before the interaction)
we have $B_1(\Delta\varphi/\ve)=0$ and
$B_2(\Delta\varphi/\ve)=1$ up to $O(\ve^N)$ 
and Eqs.~(4) describe the propagation of noninteracting shock waves.
We write 
$$
\varphi_{10}=(2u_0+u_1)t+a_1,\quad
\varphi_{20}=(2(u_0+u_1)+u_2)t+a_2,
$$
then $\psi_0(t)=\varphi_{20}-\varphi_{10}$
is the distance between the fronts of noninteracting waves.
At time $t^*$, $\psi_0(t^*)=0$, the fronts merge. 
To construct a formula that is uniform in~$t$ and represents a
weak asymptotic solution, we shall seek the phases
$\varphi_k(t,\ve)$ of shock waves in the  form 
\begin{equation}
\varphi_k(t,\ve)=\varphi_{k0}(t)+\psi_0\phi_{k1}(\tau),
\label{5}
\end{equation}
where $\tau=\psi_0(t)/\ve$ and it is assumed that
$$
\phi_{k1}(\tau)\bigg|_{\tau\to-\infty}=0,\qquad 
\frac{\phi_{k1}}{d\tau}\bigg|_{|\tau|\to\infty}=o(\tau^{-1}).
$$
Calculating the limit values of 
$\phi_{k1}(\infty)=\phi^+_{k1}$, obtain formulas 
that describe the coordinates of the fronts of shock waves 
$\varphi^+_k(t)$ after the interaction.
Substituting expressions (5) into Eq.~(4), we obtain
$$
\frac{d\varphi_{k0}}{dt}+
\frac{d\psi_{0}}{dt}\frac{d}{d\tau}[\tau\varphi_{k1}(\tau)]
=2u_0+u_1+2u_{3-k}B_k\Big(\frac{\Delta\varphi}{\ve}\Big),\quad k=1,2.
$$
We calculate the difference of these equation:
$$
\frac{d\rho}{d\tau}=F(\rho),\quad \rho=\frac{\Delta\varphi}{\ve},\quad
F(\rho)=2B_2(\rho)-1.
$$
The boundary condition for this equation has the form 
$\rho/\tau\big|_{\tau\to-\infty}\to1$.
The equation $F(\rho)=0$ has a single root $\rho_0$ 
and $B_2(\rho_0)=B_1(\rho_0)=1/2$, 
which implies that after the
interaction ($\psi_0>0$, $\tau=\psi_0/\ve\to\infty$)
the wave fronts move with the same velocity
$$
\frac{d\varphi^+_k}{dt}=2u_0+u_1+u_2,\qquad k=1,2.
$$
From~(4) for the functions $\phi_{k1}$ we obtain
$$
\phi_{k1}=(-1)^{k-1}\frac{2u_{3-k}}{(u_1+u_2)\tau}
\int^\tau_0[B_2(\rho)-1]\,d\tau'.
$$
The weak limit $u_0(x,t)$ of the weak asymptotic solution
$u_\ve(x,t)$ satisfies the classical definition of the
generalized solution (in the form of integral identity)
and the stability condition.

{\bf(b)} {\em Interaction of weak discontinuities. 
Generation and decay of shock waves}.

We again consider the Hopf equation and pose the following
initial condition
$$
u\big|_{t=0}=u^0_0+u^0_1(a_1-x)_+-u^0_1(a_2-x)_+,
$$
where  $a_1>a_2$, $z_+=z\theta(z)$, $u^0_i={\rm const}>0$
(see Fig.~1).

\begin{figure}
\centering
\includegraphics{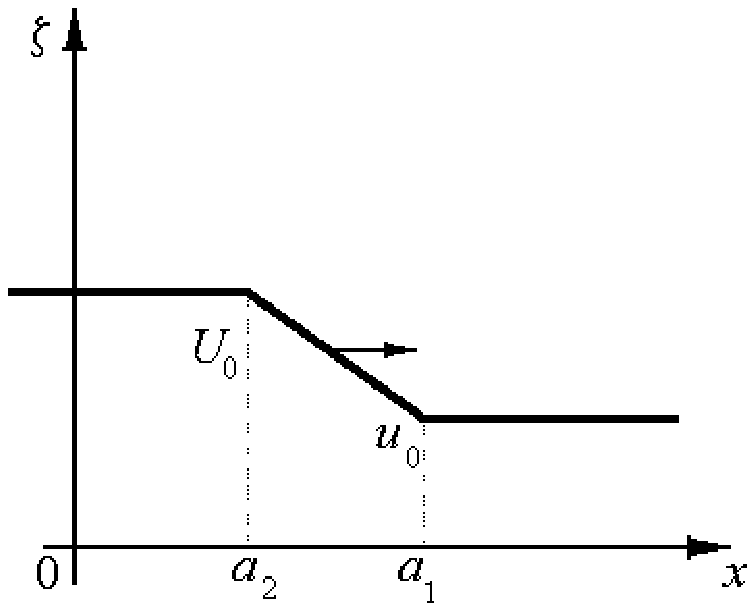}
\caption[]{}
\end{figure}

We shall seek the weak asymptotic solution in the form
\begin{eqnarray*}
u_\ve(x,t)&=&u_0+u_1(\varphi_1(t,\ve)-x)\theta_{\ve,1}(-x+\varphi_1(t,\ve))
\\
&&\qquad 
-u_2(\varphi_2(t,\ve)-x)\theta_{\ve,2}(-x+\varphi_2(t,\ve)).
\end{eqnarray*}
In this case, the equations for the functions $u_i=u_i(t,\ve)$ 
and 
$\varphi_i=\varphi_i(t,\ve)$ are derived by using a somewhat
different technique  
than that used for studying shock waves. 
Substituting the approximation of $u_\ve(x,t)$ into the equation
and taking into account the definition, 
we obtain 
\begin{eqnarray*}
&&\big(u_1(\varphi_1-x)_+\big)_t
-\big(u_2(\varphi_2-x)_+\big)_t
+\big(u^2_1(\varphi_1-x)^2_+\big)_x
+\big(u^2_2(\varphi_2-x)^2_+\big)_x\\
&&\qquad 
+2\big[u_0u_1(\varphi_1-x)_+\big]_x
-2\big[u_0u_2(\varphi_2-x)_+\big]_x\\
&&\qquad 
-2\big[u_1u_2(\varphi_1-x)(\varphi_2-x)\theta(\varphi_1-x)\big]_x
B_1(\Delta\varphi/\ve)\\
&&\qquad 
-2\big[u_1u_2(\varphi_1-x)(\varphi_2-x)\theta(\varphi_2-x)\big]_x
B_2(\Delta\varphi/\ve)=0,\quad 
\Delta\varphi=\varphi_2-\varphi_1.
\end{eqnarray*}
Let us consider the domain $\varphi_2<x\leq \varphi_1$. 
We obtain
\begin{eqnarray*}
&&u_{1t}(\varphi_1-x)+u_1\varphi_{1t}
+2\big[u_0u_1(\varphi_1-x)\big]_x
-\big[u^2_1(\varphi_1-x)^2\big]_x\\
&&\qquad 
+2u_1u_2(\varphi_1-x)B_1+2u_1u_2(\varphi_2-x)B_2=0
\end{eqnarray*}
We set $x=\varphi_1$. 
Since we have $\frac{\pa u_i}{\pa x}\equiv0$ in our example, 
we obtain 
$$
+u_1\varphi_{1t}-2u_0u_1+2u_1u_2\Delta\varphi B_1=0.
$$
Substituting this relation into the last equation, 
we arrive at the following equation for the function $u_1$:
$$
u_{1t}-2u^2_1+4u_1u_2B_1=0.
$$
In a similar way, considering the domain
$-\infty<x\leq\varphi_2$, 
we obtain the other two equations
\begin{eqnarray*}
\varphi_{2t}-2u_0+2u_1\Delta \varphi B_2(\Delta\varphi/\ve)&=&0,\\
u_{2t}+2u^2_2-4u_1u_2B_2(\Delta\varphi/\ve)&=&0,\quad
\Delta\varphi=\varphi_2-\varphi_1.
\end{eqnarray*}

Let $\Delta\varphi<0$, then, up to $O(\ve^N)$, 
we have $B_1(\Delta\varphi/\ve)=0$, 
$B_2(\Delta\varphi/\ve)=1$ and obtain the following system of
equations describing the evolution of the broken line
until it turns over:
\begin{eqnarray}
&&(\varphi_{10})'_t-2u_0=0,\qquad
(\varphi_{20})'_t-2u_0+2u_{10}(\varphi_{20}-\varphi_{10})=0,
\label{6}\\
&&(\varphi_{10})'_t-2(u_{10})^2=0,\qquad
(u_{20})'_t+2u^2_{20}-4u_{10}u_{20}=0,
\nonumber
\end{eqnarray}
Solutions of this system have the form
$$
u_{10}(t)=u_{20}(t)=u^0_1/(1-2tu^0_1),
$$
$$
\varphi_{10}=a_1+2u_0t,\qquad
\varphi_{20}=a_2+2[u^0_1(a_1-a_2)+u_0]t.
$$

We write $\psi_0=\varphi_{20}(t)-\varphi_{10}(t)$. 
At time $t=t^*$ such that $\psi_0(t^*)=0$ 
the weak discontinuities merge and a shock wave is generated.

To construct formulas that are uniform in~$t$ and describe 
the confluence of weak discontinuities 
and the generation of a shock wave, 
we seek the solution of Eqs.~(6) in the form 
$$
\varphi_k(t,\ve)=\varphi_{k0}(t)+\psi_0\phi_{k1}(\tau),\quad
\tau=\psi_0/\ve,
$$
where the functions $\phi_{k1}(\tau)=$ satisfy the same
conditions as in item~{\bf 2\,(a)}

We shall seek the functions $u_\ve(t,\ve)$ in the form
$$
u_k(t,\ve)=\psi_0(0)u^0_1/(\psi_0+\ve g_k(\tau)).
$$
Here we assume that the functions $g_k(\tau)$ behave in the same
way as the functions $\phi_{k1}(\tau)$
and take into account the relation
$$
u_{10}/u^0_1=1/(1-2tu^0_1)=\psi_0(0)/\psi_0(t),
$$
follows from the equation $\psi_{0t}+2u_{10}\psi_0=0$.
After simple calculations we see that the function $g=g_1=g_2$
satisfies the equation $\dot g+2(1-B_2(\rho))=0$ and the
function $\rho=\rho(\tau)=\Delta\varphi/\ve$ is a solution of
the boundary problem
$$
\dot\rho=1-2B_1(\rho),\qquad \rho/\tau\big|_{\tau\to-\infty}\to1.
$$
As before, the equation $1-2B_1(\rho)$ has a single root
$\rho=\rho_0$ such that $B_1(\rho_0)=B_2(\rho_0)=1/2$ and 
$\rho\to\rho_0$ as $\tau\to\infty$.
This allows us to calculate the solution for $\Delta\psi_0>0$ 
(i.e., after the interaction) or as $\tau\to\infty$.

We introduce the function $G(\tau)=\tau+g(\tau)$. 
Obviously, $\dot G=\dot\rho$,
$G/\tau\big|_{\tau\to-\infty}\to+1$, and we choose
$$
G=-\int^{\infty}_{-\infty}(1-2B_1(\rho))\,d\tau'+\rho_0.
$$
On the other hand, we can express the functions $u_i$ via the
function $G$:
$$
u_i=\frac{\psi_0(0)u^0_1}{\ve G}\stackrel{\tau\to\infty}{\to}
\frac{\psi_0(0)u^0_1}{\ve\rho_0}.
$$
We calculate the limit $(\varphi_k)^+_t$ as $\tau\to\infty$ of
the velocities of the weak discontinuities 
\begin{eqnarray*}
&&(\varphi_2)^+_t
=2u_0-\frac{2\psi_0(0)u^0_1}{\ve\rho_0}\frac12\ve\rho_0
=2u_0+(a_1-a_2)u^0_1,\\
&&(\varphi_1)^+_t
=2u_0-\frac{2\psi_0(0)u^0_1}{\ve\rho_0}\frac12\ve\rho_0
=2u_0+(a_1-a_2)u^0_1,
\end{eqnarray*}
which coincides with the velocity of the shock wave
$$
U(x,t)=u_0+(a_1-a_2)u^0_1\theta(-x+\varphi^+(t)),
$$
where $\varphi^+=(\varphi_2^+)_t=(\varphi^-_1)_t$.
By using the explicit formula for the solution $u_\ve(x,t)$,
we can easily show that 
$$
w-\lim_{\ve\to0}u_\ve(x,t)=U(x,t), \qquad t>t^*.
$$
To this end, we rewrite the above-constructed solution
$u_\ve(x,t)$ in the form 
$$
u_\ve(x,t)=u_0
+u_1(\varphi_1-\varphi_2)\theta_{\ve,1}(\varphi_1-x)
+u_1(x-\varphi_2)
\big[\theta_{\ve,2}(\varphi_2-x)-\theta_{\ve,1}(\varphi_1-x)\big].
$$
Consider the second term. We have
$$
u_1(\varphi_1-\varphi_2)
=\frac{\psi_0(0)u^0_1\rho}{G}
=\psi_0(0)u^0_1=(a_1-a_2)u^0_1
\stackrel{\mbox{def}}{=}U_0.
$$
Since $\varphi_1\big|_{t>t^*}\simeq\varphi^+$, 
the first two terms pass into the shock wave $U(x,t)$ for
$t>t^*$.  
Consider the last term
$$
u_1(x-\varphi_2)
\big[\theta_{\ve,2}(\varphi_2-x)-\theta_{\ve,1}(\varphi_1-x)\big]
=u_1(x-\varphi_2)\bigg[
\frac{\theta_{\ve,2}(\varphi_2-x)-\theta_{\ve,1}(\varphi_1-x)}
{\varphi_1-\varphi_2}\bigg]
$$

As was already shown, the coefficient of the expression in
braces is a constant. 
The expression in square brackets is an approximation of the
$\delta$-function at the point $\varphi_2$.
Hence the entire expression in braces is small 
(in a uniform metric) as $\ve\to0$.

We study the problem in which a shock wave is generated by a
special (piecewise linear) initial condition. 
The case of a general smooth initial functions can be treated
similarly. 
Here we need to consider a family of linear interpolations 
of this initial condition and to use the above technique 
on segments of the broken line. 

To study this problem in more detail, we note that we have
considered only one possibility of evolution of the broken line,
namely, formation of a step. 
Another mechanism of evolution is as follows:  
segments of the broken line are added to the step that has
already been formed.  
This is the {\em confluence of a weak discontinuity and a shock
wave}. 

Now we again consider the Hopf equation in order to study this
mechanism. 
The initial condition corresponding to this type of interaction
has the form
$$
u\big|_{t=0}=u^0_0\theta(a^0_1-x)
+u^0_1(a_1-x)\theta(a_1-x)
-u^0_1(a_2-x)\theta(a_2-x),
$$
where $u^0_0,\,u^0_1$ are positive constants and 
$a_1>a_2$ (see Fig.~2).

\begin{figure}
\centering
\includegraphics{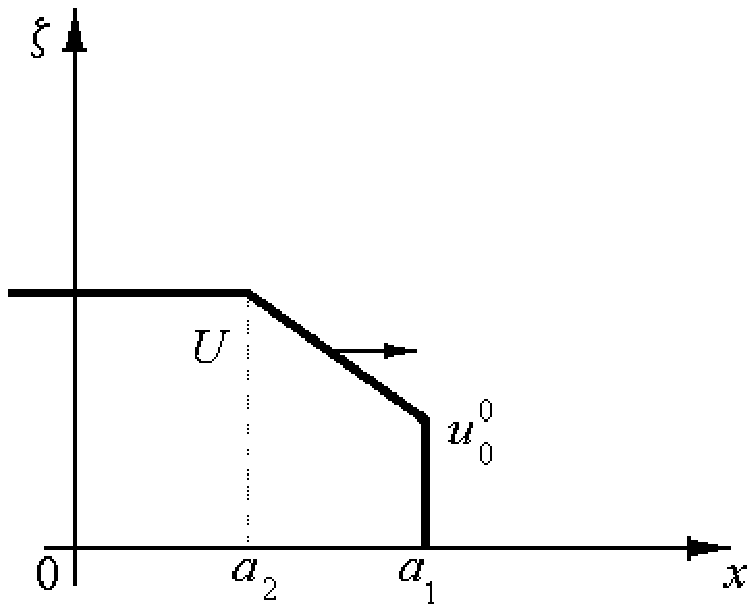}
\caption[]{}
\end{figure}

Just as before, we construct the approximation of the solution
in the form 
$$
u_\ve(x,t)=u_0\theta_{\ve,1}(\varphi_1-x)
+u_1(\varphi_1-x)\theta_{\ve,1}(\varphi_1-x)
-u_1(\varphi_2-x)\theta_{\ve,2}(\varphi_2-x),
$$
where $u_i=u_i(t,\ve)$, $\varphi_i=\varphi_i(t,\ve)$.
Substituting this expression into the Hopf equation, we obtain
the system of equations (cf. above)
\begin{eqnarray}
\varphi_{1t}-u_0+2u_1\psi B_1&=&0,\nonumber\\
u_{1t}-2u^2_1+4u^2_1u_2B_1&=&0,\\
\varphi_{2t}-2u_0B_2+2u_1\psi B_2&=&0,\nonumber\\
u_{0t}-u_0u_1(1-2B_1)&=&0,\nonumber
\end{eqnarray}
where $B_i=B_i(\Delta\varphi/\ve)$ are the functions derived
above, $\Delta\varphi=\varphi_2-\varphi_1$.
Before the interaction, we have $\varphi_2<\varphi_1$, 
$\Delta \varphi/\ve\sim-\infty$, and $B_1=0$, $B_2=1$ 
with arbitrary accuracy in $\ve$.
Denoting by $\varphi_{10},\,u_{10},\,\varphi_{20},\,u_{00}$ the
solution of system (7) with $B_1=0$, $B_2=1$, 
we obtain the following system of equations for these functions: 
\begin{eqnarray*}
\varphi_{10t}&=&u_{00},\\
u_{10t}&=&2u^2_{10},\\
u_{00t}&=&u_{00}u_{10},\\
\varphi_{20t}&=&2(u_{00}-u_{10}\psi_0),\qquad
\psi_0=\varphi_{20}-\varphi_{10}.
\end{eqnarray*}
It is easy to find the solution of this system:
\begin{eqnarray*}
u_{10}&=&\frac{u^0_1}{1-2u^0_1t},\qquad 
u_{00}\,=\,\frac{u^0_0}{(1-2u^0_1t)^{1/2}},\\
\varphi_{20}&=&a_2+2Ut,\qquad 
\varphi_{10}\,=\,a_1+\int^t_0 u_{00}\,dt,\\
\psi_0&=&\frac1{u^0_1}\Big[(\psi^0_0u^0_1-u^0_0)(1-2u^1_0t)
-u^0_0\sqrt{1-2u^1_0t}\Big],\\
U&=& u^0_0+u^0_1(a_1-a_2).
\end{eqnarray*}
  
One can easily see that the function $\psi_0(t)$ vanishes at the
two points $t_1=1/2u^1_0$ and $t^*$ such that 
$$
\sqrt{1-2u^1_0t^*}=\frac{u^0_0}{U}.
$$
Obviously, $t^*<t_1$ and the free singularities supports 
$x=\varphi_{10}$ and $x=\varphi_{20}$ merge at $t=t^*$.
In this case we have
$$
u_{00}(t^*)=u^*_{00}\equiv U,\qquad
u_{10}(t^*)=\frac{U^2}{(u^0_0)^2}u^0_1<\infty.
$$
Thus in this example  the mechanism of formation of a new shock 
wave consists not in turning over the inclined segment of
the broken line, as in the preceding example, but in the
disappearance of this inclined segment due to increasing 
vertical segment. 

Subtracting the first equation from the third equation in (7),  
we obtain the following equation for the function $\psi$:
$$
\psi_t=(u_0-2\psi u_1)(1-2B_1(\Delta\varphi/\ve))
$$
or, denoting
$\rho=\Delta\varphi/\ve=(\psi_0+\psi_0\psi_1(\tau))/\ve$, $\tau=\psi_0/\ve$,
$$
\psi'_0\dot\rho=[u_0-2\psi u_1](1-2B_1(\rho)),\qquad 
\frac{\rho}{\tau}\bigg|_{\tau\to-\infty}\to1.
$$
Note that we can use the formula for $\psi_0$
(and for the functions $u_{00}$ and $u_{10}$) 
only for $t\in[0,t^*+\delta]$, where $\delta>0$ is any number
such that $\delta<t_1-t^*$. 

To obtain formulas that are global in~$t$, we need to choose a
number $\delta$ and continue the functions $u_{00}$, 
$u_{10}$, and $\psi_0$ smoothly 
to the time $t\geq t^*+\delta$ so that the sign is preserved.
Calculating the coefficient of $u_0-2\psi u_1$, one can easily
see that $\rho>0$ for $t<t^*$.
Hence there exists a solution $\rho\to\rho_0$, where $\rho_0$ 
is a root of the equation 
$$
B_1(\rho)=1/2.
$$
Let us consider the system of equations for the functions 
$u_0$ and $u_1$. 
By the change $u_0=u^0_0\sqrt{u_1/u^0_1}$,
this system can be reduced to the single equation for $u_1$:
$$
u_{1t}-2u^2_1(1-2B_1)=0.
$$
Its solution has the form 
$$
u_{1}(t,\ve)=\frac{u^0_1}{1-2u^0_1\int^t_0[1-2B_1(\rho(\tau))]\,dt'}.
$$
Clearly, we have $u_1(t^*,\ve)\leq u_{10}(t^*)$ 
(since $\int^{t^*}_0(1-2B_1)\,dt'\leq t^*$).
On the other hand, we have $t>t^*$ for $\rho\to\rho_0$.
Therefore, $\psi_1(\tau)\to-1$ as $\tau\to\infty$ 
and hence 
$(\Delta\varphi)\to0$ as $\tau\to\infty$ 
(i.e., for $t>t^*$, $\ve\to0$).
This implies that for $t>t^*$ we have
$$
u_1(t,\ve)=u_{10}(t^*)+o(1),\qquad \ve\to0.
$$
We represent the above-constructed solution in the form 
\begin{eqnarray*}
u_\ve(x,t)&=& U\theta_{\ve,1}(\varphi_1-x)
+(u_0-U)\theta_{\ve,2}(\varphi_2-x)\\
&&+[\theta_{\ve,1}(\varphi_1-x)-\theta_{\ve,2}(\varphi_2-x)]
(\varphi_2-x)u_1.
\end{eqnarray*}
Obviously, for $t>t^*$, $\ve\to0$, 
the first term approximates the shock wave
$$
u=U\theta(a_1+Ut-x)
$$
and in this case the second term vanishes since $u_0-U\to0$
and the third term vanishes since
$\varphi_1-\varphi_2=\Delta\varphi\to0$. 
Recall that $\psi_{0t}=u_{00}-2\psi_0u_{10}$ for $t<t^*$.
In view of the equations, 
we can continue the function $\psi_{0t}$ for $t>t^*$ in the form
$\psi_{0t}=U$. 
In this case the function $u_0-2\psi u_1$ is continuous
uniformly in $\ve$ for $t=t^*$ and we can show that the function
$\rho$ is a solution of boundary value problem 
$$
\dot\rho=(1-2B_1(\rho)),\qquad 
\frac{\rho}{\tau}\bigg|_{\tau\to-\infty}\to1.
$$
The system of equations determining the weak asymptotic solution
in this case also splits into separate equations.

Now we briefly consider the problem of {\em decay of nonstable
shock waves}. 

One can easily see that by setting
$v_{T,\ve}(x,t)=u_\ve(x,T-t)$, $T>t^*$, we obtain a
$T$-dependent family of weak asymptotic solutions of the
equation $v_t-(v^2)_x=0$. 

For $t=0$ the solutions of this family are shock waves 
(unstable for this new equation).
The weak limit of these solutions for $0\leq t<T-t^*$
is a shock wave, 
and for $t>T-t^*$ is a broken line consisting of two moving
weak discontinuities into which the unstable shock wave splits
at time $t_*(T)=T-t^*$ (which is not unique).

{\bf(c)} {\em Interaction of shock waves in the multidimensional
case}. 
Let us consider the two-dimensional nonlinear equation 
arising in the reservoir problem
$$
L[u]=\frac{\pa u}{\pa t}+A_1\frac{\pa u^2}{\pa x_1}
+A_2\frac{\pa u^2}{\pa x_2}=0.
$$

The above approach can be easily generalized to the case of an
arbitrary dimension if the codimension of the front of the
nonlinear wave is $=1$.
We assume that $A_1\ne A_2$ are positive constants.

We choose the initial conditions as 
$$
u\big|_{t=0}=u_0+u_1\theta(t+\psi_1(x))+u_2\theta(t+\psi_1(x)),
$$
where $x=(x_1,x_2)$, $u_i$ are positive constants, 
and $\psi_i(x)$ are the desired functions. 
We write $\Gamma^0_i=\{x,\psi_i(x)=0\}$.

Clearly, the curves $\Gamma^0_i$ are given initial positions of
the fronts of two shock waves whose sum is just the initial
condition. 
In addition, we assume that the curves $\Gamma^0_i$ 
are transversal to the vector field 
$\langle\vec A,\nabla\rangle$, $\vec A=(A_1,A_2)$ and $\Gamma^0_2$
is the cross-section of the (trivial) fibration over $\Gamma^0_1$
whose fibers are straight lines parallel to the vector~$\vec A$.
In addition, we assume that the motion from the points 
of $\Gamma^0_2$ to the points of $\Gamma^0_1$ is in the
direction of the vector~$\vec A$. 
In this case the fact 
that $u_1,u_2$ are positive constants is a sufficient condition
of stability. 

If the functions $\psi_i(x)$ are known, then the curves 
(level surfaces) $\Gamma^t_i\{x,t+\psi_i(x)=0\}$
determine the fronts of shock waves at time~$t$.

Acting as before (see {\bf2\,(a)}), we substitute the
approximation  
$$
u_\ve(x,t)=u_0+u_1\theta_{\ve,1}(t+\psi_1(x,\ve))
+u_2\theta_{\ve,2}(t+\psi_2(x,\ve))
$$
into the equation and calculate the weak asymptotics of
$L[u_\ve]$. 
We obtain 
\begin{eqnarray*}
L[u_\ve]&=&\delta_{\Gamma^t_1}\Big[1
+\langle\vec A,\nabla\psi_1\rangle(u_1+2u_0+2u_2B_1(\Delta\psi/\ve))\Big]\\
&&+\delta_{\Gamma^t_2}\Big[1
+\langle\vec A,\nabla\psi_2\rangle(2u_0+u_2+2u_1B_2(\Delta\psi/\ve))\Big]
+O_{\cD'}(\ve),\\
\Delta\psi&=&\psi_2-\psi_1.
\end{eqnarray*}
Here the functions $B_1$ and $B_2$ are the same as in item~{\bf2\,(a)}.
Formulas for the weak asymptotics in the multidimensional case
are carried out in the same way as in the one-dimensional case

Roughly speaking, the (two-dimensional) integral becomes an
iterated integral over the surface $\Gamma^t_i$ 
and over the normal to this surface. 
The asymptotics of the integral along the normal is calculated
in the same way as in the one-dimensional case, see [1].

It follows from our assumptions that the inequality 
$\Delta\psi<0$ holds for sufficiently small positive~$t$.
Hence we have $\Delta\psi/\ve\sim-\infty$ and for small $t$ we
obtain the following equations describing the system of
noninteracting fronts:
\begin{equation}
1+\langle \vec A,\nabla\psi_1\rangle(u_1+2u_0)=0,\qquad
1+\langle \vec A,\nabla\psi_2\rangle(u_2+2u_1+2u_0)=0.
\end{equation}

Clearly, these equations determine the (limit) functions
$\psi_{k0}$ if the curves $\Gamma^0_i$ on which they vanish are
given. 

Dividing these equations by $|\nabla\psi_i|$ and taking into
account the fact that, in view of our formulas, the waves travel
in the direction of decreasing functions $\psi_{k0}(x)$,
we can rewrite the last system as   
$$
V^{(1)}_{n_1}=\langle\vec A,\vec n_1\rangle(u_1+2u_0),\qquad
V^{(2)}_{n_2}=\langle\vec A,\vec n_2\rangle(u_2+2u_1+2u_0),
$$
where $n_i$ is the normal (at a point) to $\Gamma^t_i$,  
$V^{(i)}_{n_i}$ is the normal velocity of this point.
Clearly, 
the velocities of the points of the curve $\Gamma^t_2$ 
are larger than the velocities of the points 
of the curve $\Gamma^t_2$ along the trajectories 
of the field $\langle\vec A,\nabla\rangle$, 
but the distance between the curves $\Gamma^t_i$ along 
the trajectory depends, in general, on the point.

Therefore, since the shape of these curves is rather arbitrary,
there may be no complete confluence of these curves at their
contact. 
A new shock wave with summary amplitude $u_1+u_2$ is generated at
the points of contact. This shock wave travels with its new
velocity, and the solution may be of a rather complicated
structure. 
To describe this wave uniformly in time, 
we shall seek the solution of the system
\begin{eqnarray}
1+\langle \vec A,\nabla\psi_1\rangle(u_1+2u_0+2u_2B_1(\Delta\psi/\ve))=0,
\label{8}\\
1+\langle \vec A,\nabla\psi_2\rangle(u_2+2u_0+2u_1B_2(\Delta\psi/\ve))=0
\nonumber
\end{eqnarray}
in the form 
\begin{equation}
\psi_{k}(x,\ve)=\psi_{k0}(x)+\phi_0(x)\psi_{k1}(\phi_0/\ve),
\label{9}
\end{equation}
where $\phi_0=\psi_{20}(x)-\psi_{10}(x)$. Note that, 
in view of our assumptions on the geometry, 
instead of the coordinates $(x_1,x_2)$, 
we can introduce the coordinates $(s,\xi)$, 
where $s$ are the coordinates on $\Gamma^0_2$ and $\xi$ is a
parameter on the trajectories of the vector field 
$\langle\vec A,\nabla\rangle$.
 
Hence we, in fact, ``calculate the distance'' between the
curves $\Gamma^+_i$  (i.e., the differences $\Delta\psi$,
$\phi_0$) along the trajectories of the field  
$\langle\vec A,\nabla\rangle$.

Preserving, instead of $\frac{d}{d\xi}$, 
the notation $\langle\vec A,\nabla\rangle$,
substituting (9) into (8), and taking into account (10), we
obtain 
\begin{eqnarray}
&&2\langle\vec A,\nabla\psi_{10}\rangle u_2B_1
+\langle\vec A,\nabla\phi_0\rangle
\frac{d}{d\tau}(\tau\psi_{11})
[U_0+2u_2(B_1-1/2)]=0,
\label{10}\\
&&-2u_1\langle\vec A,\nabla\psi_{20}\rangle
+2\langle\vec A,\nabla\psi_{20}\rangle u_1B_2
\nonumber\\
&&\qquad 
+\langle\vec A,\nabla\phi_0\rangle
\frac{d}{d\tau}(\tau\psi_{21})
[U_0+2u_1(B_2-1/2)]=0.
\nonumber
\end{eqnarray}
Here $U_0=u_1+u_2+2u_0$, $\tau=\phi_0/\ve$.  

The further is similar to that in the one-dimensional case. 
Its first stage is to obtain an equation for the function 
$\rho=\Delta\psi/\ve=(\phi_0/\ve)
(1+\psi_{21}(\tau)-\psi_{11}(\tau))\stackrel{\rm def}{=}
\tau(1+\phi_1(\tau))$.
Next, we calculate the limits of the functions 
$B_k(\rho)$ as $\tau\to\infty$ (after the interaction)
and find equations for the limit functions
$\psi^+_k$ as well as equations for $\psi_{k1}(\tau)$. 

Subtracting the first equation in (10) form the second one
and carrying out several calculations, 
we obtain the desired equation for $\rho$:
$$
\dot\rho=1-\frac{1-B_2(\rho)}{u_1+u_2}
\bigg[\frac{2u_2U_2}{U_0-2u_2(B_2-1/2)}
+\frac{2u_1U_1}{U_1+2u_1(B_2-1/2)}\bigg],
\quad
\frac{\rho}{\tau}\bigg|_{\tau\to-\infty}\!\!\!\to1.
$$

One can show that the right-hand side of this equation 
(that differs, as one can see, from that in the similar equation
in the one-dimensional case)
also has a single root $\rho_0$ and 
$B_2(\rho_0)=1/2$ (and hence $B_1(\rho_0)=1/2$).

Hence it follows from Eq.~(8) that, for the same values of~$s$
for which a point of the curve $\Gamma^+_1$ ``outruns''
the curve $\Gamma^+_1$,
we have 
$$
1+\langle \vec A,\nabla\psi^+_k\rangle(u_1+u_2+2u_0)=0,\qquad
k=1,2.
$$
This implies $\psi^+_1=\psi^+_2$ and for a given~$s$, 
after the confluence of the curves, 
a wave with summary amplitude $u_1+u_2$ travels 
in the direction of $\vec A$.
Thus, for a fixed~$s$, 
the dynamics of interaction in the direction of $\vec A$
is similar to that in the one-dimensional case. 

Here we do not write out the equations for $\psi_{k1}$.
They can be obtained in the same way as the similar equations
in the one-dimensional case. 

\section{Conclusion}

The problem of shock wave interaction in the one-dimensional
case is presented in [2, 3, 5] not only for the Hopf equations but 
also for equations with sufficiently general nonlinearity. 
The formulas from Sec.~1 are derived there in more detail.

Similarly, in the multidimensional case one can easily
generalize our construction to the case of more general 
nonlinearities, variable coefficients and amplitudes.

For reasons of space, here we do not consider the problem of
constructing definitions of weak solutions. This problem is
discussed in [1, 4, 5]. 
In particular, in [5] a definition of a weak solution is
constructed for KdV type equations admitting the zero dispersion
limit for soliton type solutions.

\end{document}